\documentclass[10pt,conference]{IEEEtran}
\usepackage{latexsym}
\usepackage{amsmath}
\usepackage{amssymb}
\usepackage{bm}
\usepackage[dvipdfm]{graphicx}
\usepackage{latexsym}
\usepackage{amsmath}
\usepackage{amssymb}
\usepackage{bm}
\usepackage{amsmath,amssymb}
\usepackage{graphicx}

\newcommand{\gb}[2]{\left[\!\!\!\begin{array}{c}#1\\#2\end{array}\!\!\!\right]}

\newcommand{\R}{{\mathbb R}}

\newcommand{\GL}{\mathrm{GL}}
\newcommand{\GF}{\mathrm{GF}}

\newcommand{\tentative}{\noindent{{\underline{\tt tentative decision}}} }

\newcommand{\ctov}{\noindent{{\underline{\tt check to variable output}}} }
\newcommand{\vtoc}{\noindent{{\underline{\tt variable to check output}}} }

\newcommand{\initialization}{\noindent{{\underline{\tt initialization}}} }

      \title{Fountain Codes with Multiplicatively Repeated Non-Binary LDPC Codes}
      \author{
      Kenta KASAI and  Kohichi SAKANIWA, \\
Tokyo Institute of Technology, 152-8550 Tokyo, JAPAN,  \{kenta, sakaniwa\}@comm.ss.titech.ac.jp
      }  
\begin{document}
      \maketitle
      \begin{abstract}
       We study fountain codes transmitted over the binary-input symmetric-output channel. 
       For channels with small capacity, receivers needs to collects many channel outputs to recover 
       information bits. 
       Since a collected channel output yields a check node in the decoding Tanner graph, 
       the channel with small capacity leads to large decoding complexity. 
       In this paper, we introduce a novel fountain coding scheme with non-binary LDPC codes.
       The decoding complexity of the proposed fountain code does not depend on the channel. 
       Numerical experiments show that the proposed codes exhibit better performance than conventional fountain codes, especially for small number of information bits. 
      \end{abstract}
 \section{Introduction}
 \label{155959_8Apr10}
Fountain codes are a class of erasure-recovering or error-correcting codes which produce limitless sequence of encoded bits 
from $k$ information bits so that receivers can recover the $k$ information bits from any $(1+\epsilon)k/C$ encoded bits, where $C$ is 
the channel capacity and $\epsilon$ is referred to as {\itshape overhead}. 
The name is after water fountains which endlessly produce water drops to entertain people.  
 Designing fountain codes with small overhead is desirable. 
LT codes \cite{1181950} and  Raptor codes \cite{raptor} are fountain codes which 
achieves vanishing overhead $\epsilon\to 0$ in the limit of large information size 
over the channel with $C=1$, i.e., the binary erasure channel (BEC). 
By a nice analogy between the BEC and the packet erasure channel, fountain codes successfully adopted by several industry standards. 
 
In \cite{1624639}, Etesami et al. investigated Raptor codes used over the memoryless binary-input output-symmetric (MBIOS) channels. 
And they 
showed that over the AWGN channels with capacity $C\ge 0.49$, 
Raptor codes achieve overhead $\epsilon\le 0.08$ at BER $10^{-7}$ with information size $k=65536$. 
A Raptor code can be viewed as concatenation of an outer high-rate LDPC code and infinitely many single parity-check codes of length $d$, 
where $d$ is chosen randomly with probability $\Omega_d$ for $d\ge 1$. 
In \cite{1702289}, Venkiah et al. proposed a joint decoding of the concatenated codes and an optimization method for output degree distributions $\Omega(x):=\sum_{d\ge 1}\Omega_dx^d$  and showed that the optimized codes outperform the conventional ones.

    The problems for constructing fountain codes used for general channels with finite inputs  are summarized as follows. 
    \begin{itemize}
    \item {\bfseries Problem 1}: The output degree distribution $\Omega(x)$ needs to be optimized for each $k$. And large check node degree $d$ leads to the large encoding and decoding complexity.
    \item {\bfseries Problem 2}: The number of check nodes in the inner codes is given by $(1+\epsilon)k/C$. This increases as the channel capacity $C$ decreases.  
	  Since check node computation is dominant in decoding, the decoding complexity is high for small $C$. 
    \item {\bfseries Problem 3}: Large size of information and vanishing overhead are often considered. 
	  This leads to large size of memory devices and transmission latency.   
   \end{itemize}
    In this paper, we will propose a novel fountain coding scheme which is free of those drawbacks. 

    In this paper, we consider non-binary LDPC codes defined by sparse parity-check matrices over $\GF(2^m)$ for $2^m>2$. 
    Non-binary LDPC codes are invented by Gallager \cite{gallager_LDPC}  and, Davey and MacKay \cite{DaveyMacKayGFq} found non-binary LDPC codes can outperform binary ones. 
    Non-binary LDPC codes have captured much attention recently due to their decoding performance. 

    It is known that the irregularity of Tanner graphs helps improve the decoding performance of binary LDPC codes. 
    While, it is not the case for the non-binary LDPC codes. 
    Interestingly, the $(2, d_\mathrm{c})$-regular non-binary LDPC codes over $\GF(2^m)$ are empirically known \cite{4641893} as the best performing codes for $2^m\ge 64$, 
    especially for short code length. 
    This means that, for designing non-binary LDPC codes, one does not need to optimize the degree distributions of Tanner graphs, since $(2, d_\mathrm{c})$-regular non-binary LDPC codes are best. 
    Furthermore, the sparsity of $(2,d_\mathrm{c})$-regular Tanner graph makes efficient decoding possible. 


    

    \section{Fountain Coding with Multiplicatively Repeated Non-Binary LDPC Codes}
    \label{223813_17Feb10}
    In this section we explain a new fountain coding scheme. 
    The new coding scheme uses a non-binary LDPC code as a pre-code. 

    In \cite{submitted_nb_lr}, the authors presented low-rate non-binary codes. 
    The code is a concatenation of $(2,3)$-regular non-binary LDPC code and inner multiplicative repetition codes. 
    In general, low-rate LDPC codes have many check nodes and suffer from the high decoding complexity than hight rate codes. 
    One of the remarkable features of the code is that the decoding complexity does not depend on the coding rate. 
    The code exhibits excellent decoding performance for small code length and is rate-compatible. 
    We will use the low-rate code \cite{submitted_nb_lr} with vanishing rate  as a fountain code. 

    We fix a Galois field $\GF(2^m)$ with a primitive element $\alpha$ and its primitive polynomial $\pi(x)$. 
    Once the primitive element is fixed, one can represent each symbol in the Galois field as a binary sequence of length $m$ \cite{macwilliams77}. 
    For example, with a primitive element $\alpha\in\GF(2^3)$ such that $\pi(\alpha)=\alpha^3+\alpha+1=0$, each symbol is represented as
    $0=(0,0,0)$, $1=(1,0,0)$, $\alpha=(0,1,0)$, $\alpha^2=(0,0,1)$, 
    $\alpha^3=(1,1,0)$, $\alpha^4=(0,1,1)$, $\alpha^5=(1,1,1)$ and $\alpha^6=(1,0,1)$. 
    In this setting,   $k$ information bits can be  represented as 
    $k/m$ symbol sequence $(x_1,\dotsc,x_{k/m})\in\GF(2^m)^{k/m}$.
    Note that what corresponds to a packet used in the typical fountain coding system is not the sequence but  
    each bit in symbols, i.e.,  $x_i$ for $i\ge 1$. 
    We refer to elements in $\GF(2^m)$ as symbols for $m\ge 2$ and bits for $m=1$. 

    A non-binary LDPC code $\mathcal{C}$ over $\GF(2^m)$ is defined by the null space of a sparse $M\times N$ parity-check matrix $H=\{h_{i,j}\}$ defined over $\GF(2^m)$. 
    \begin{align*}
    \mathcal{C}&=\{x\in \GF(2^m)^N\mid Hx^T=0\in\GF(2^m)^M\}
   \end{align*}
    The $c$-th parity-check equation for $c=1,\dotsc,M$ is written as
    \begin{align*}
    h_{c,1}x_1+\cdots+h_{c,N}x_N=0\in\GF(2^m),
   \end{align*}
    where $h_{c,1},\dotsc, h_{c,N}\in \GF(2^m)$ and $x_1,\dotsc, x_N\in\GF(2^m)$. 

    The binary LDPC codes are represented by Tanner graphs with variable and check nodes \cite[pp.~75]{mct}. 
    The non-binary LDPC codes, in this paper, are also represented by bipartite graphs with variable nodes and check nodes, which are also referred to as Tanner graphs. 
    For a given sparse parity-check matrix $H=\{h_{cv}\}$ over $\GF(2^m)$,  the graph is defined as follows. 
    The $v$-th variable node and $c$-th check node are connected if  $h_{cv}\neq 0$. 
    By $v=1,\dotsc,N$ and $c=1,\dotsc,M$, we also denote the $v$-th variable node and $c$-th check node, respectively. 

    A non-binary LDPC code with a parity-check matrix over $\GF(2^m)$ is called $(d_\mathrm{v}, d_\mathrm{c})$-regular if 
    all the columns and all the rows of the parity-check matrix have  weight  $d_\mathrm{v}$ and $d_\mathrm{c}$, respectively,  
    or equivalently all the variable and check nodes have degree $d_\mathrm{v}$ and $d_\mathrm{c}$, respectively.
   \begin{figure}[t!]
    \includegraphics[scale=1.1]{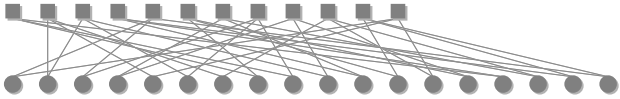}
    \vspace{-23mm}
    \caption{An example of a pre-code $\mathcal{C}_1$.  A  non-binary (2,3)-regular LDPC code of rate 1/3 over $\GF(2^m)$. 
    Each variable node represents a symbol in $\GF(2^m)$.
    Each check node represents a parity-check equation over $\GF(2^m)$.
    The code length is 18 symbols in $\GF(2^m)$ or equivalently $18m$ bits.  
    }
    \label{203721_13Jan10}
   \end{figure}
    Let $\mathcal{C}_1$ be a $(2,3)$-regular LDPC pre-code defined over $\GF(2^m)$ of length $N$ symbols or equivalently $mN$ bits and  of rate $1/3$.
    It can be seen that $N=3k/m$. 
    The pre-code $\mathcal{C}_1$ has a $2N/3\times N$  sparse parity-check matrix $H=\{h_{i, j}\}$ over $\GF(2^m)$. The matrix $H$ has row weight 3 and column weight 2. 
    Fig.~\ref{203721_13Jan10} shows the Tanner graph of $\mathcal{C}_1$ of length $N=$18 symbols. 
    It can be shown that $(2,d_\mathrm{c})$-regular non-binary LDPC codes is linear-encodable by using a non-singular zig-zag subgraph. 

    We define a new fountain code $\mathcal{C}_{\infty}:\GF(2)^k\to \GF(2)^\infty$ by giving the encoding procedure as follows. 
\begin{enumerate}
\item     First, map the $k$ information bits to $k/m$ information $\GF(2^m)$-symbols. 
\item     By the pre-code $\mathcal{C}_1$, encode the $k/m$ information symbols to $N$ symbols $x_1,\dotsc,x_N\in\GF(2^m)$ . 
\item    Repeat the followings endlessly from $i=1$ to $\infty$. 
\begin{enumerate}
 \item     Pick randomly $v_i\in [1,N]$,  $w_i\in[1,m]$ and $h_i\in\GF(2^m)\backslash\{0\}$. 
 \item    Transmit $w_i$-th bit of $h_ix_{v_i}\in\GF(2^m)$. 
\end{enumerate}
\end{enumerate}

    The proposed fountain code $\mathcal{C}_{\infty}$ can be viewed as  a non-binary Raptor code with 
    a non-binary (2,3)-regular LDPC pre-code and 
    an output degree distribution $\Omega(x)=x$ \cite{raptor}. 
    Note that $\Omega(x)=x$ does not mean simple repetition of bits but multiplicative repetition of 
    symbols in $\GF(2^m)$ for the proposed non-binary setting. 
    \section{Decoding Scheme}
    \label{223813_17Feb10}
    We assume that transmission takes place over the MBIOS channel. 
    Specifically, the channel is specified by the transition probability $P(\cdot|\cdot)$ such that
     $P(y|x)=\Pr({Y}={y}|{X}={x})$
    where $X$ and $Y$ are the random variable of an input bit $x$ and the channel output $y$, respectively. 
    And we assume that the information bits are chosen with uniform probability. 

    The most important feature of the fountain coding system is that 
    the decoder does not receive all the channel output but collects $n$ channel outputs. 
    The decoder recovers the $k$ information bits from the $n$ collected channel outputs.  
    The {\itshape overhead} $\epsilon$ is defined \cite{1624639,1702289} by 
    \begin{align*}
     \epsilon=C/R-1, \quad R=k/n,
    \end{align*}
    where $C$ is the channel capacity.
    Then, the decoder has $n=(1+\epsilon)k/C$ collected channel outputs. 
    Note that, in the original setting of fountain codes as in \cite{1181950},\cite{raptor}, 
    the capacity is set $C=1$, i.e., all the collected bits are uncorrupted. 
    The aim of the fountain coding in this paper is to reliably recover the information bits with small overhead. 
    The overhead $\epsilon=0$ implies that the information bits are transmitted at rate $R=C$, which is our extreme aim.  
    With infinitely many information bits, Raptor codes can achieve $\epsilon=0$ for the channel with $C=1$, i.e., the BEC. 
    And Raptor codes optimized for the BEC exhibit a quite good performance for large information bits with $k=65536$. 
    However, for both the BEC and the general MBIOS channels with $C<1$, Raptor codes exhibit high error floors \cite{1624639}, \cite{VPD08}, \cite{Ven08}, \cite{1702289} for small information bits with $k\sim 1024$. 

    For the $i$-th transmitting bit, the sender picked randomly $v_i\in [1,N]$,  $w_i\in[1,m]$ and $h_i\in\GF(2^m)\backslash\{0\}$
    and transmitted $w_i$-th bit of $h_ix_{v_i}\in\GF(2^m)$. 
    Let $I$ be the set of transmitting indices that the receiver collected. It follows $\#I=n$. 
    In other words, for $i\in I$, the receiver collects $y_i$ that is the corrupted version of the $i$-th transmitted bits. 
    We assume that the decoder knows not only $y_i$ but also the indices $v_i, w_i$ and the multiplicative coefficients $h_i$ for $i\in I$. 
    In practice, this is realized by embedding the indices in the header of packets or synchronization 
    between the sender and the receivers \cite{raptor}. 

   \begin{figure}
    \includegraphics[scale=1.1]{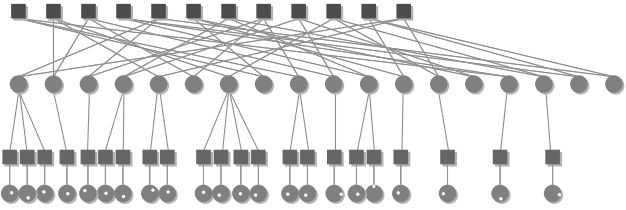}
    \vspace{-0.9cm}
    \caption{An example of a Tanner graph used for decoding. 
    Some variable nodes are of degree one. The variable nodes of degree one are corresponding to the transmitted symbols whose channel outputs are collected by the decoder. 
    White dots represent bits corresponding to the received channel outputs. 
    It can be seen that the decoder collected 22 channel outputs for this example. 
    }
    \label{175738_8Jan10}
   \end{figure}

    The proposed code $\mathcal{C}_{\infty}$ can be decoded by the sum-product (SP) decoding algorithm on the Tanner graphs. 
    The SP decoder for the non-binary LDPC codes exchanges probability vectors in $\in\R^{2^m}$, 
    called {\it messages}, between variable nodes and check nodes \cite{RaU05/LTHC}. 
    An example of the Tanner graph used by the decoder is shown in Fig.~\ref{175738_8Jan10}. 
    The variable nodes of degree one with  white dots in Fig.~\ref{175738_8Jan10} represent collected channel outputs. 
    If the SP decoding algorithm is immediately applied  to the proposed codes, 
    all the variable nodes and check nodes, including the variable nodes of those multiplicative repetition symbols, are activated, i.e.~exchage the messages. 
    However, the  messages reached at the variable nodes of degree one do not change messages that sent back from the nodes. 
    Therefore,  after the initialization, the decoder does not need to pass the messages all the way to those variable nodes of degree 1 and their adjacent check nodes of degree 2. 
    Consequently, the decoder uses only the Tanner graph of the pre-code $\mathcal{C}_1$. 
    It follows that the complexity of the decoding algorithm does not depend on the number $n$ of collected channel outputs
    and the channel capacity $C$. In contrast the decoding complexity of 
    the conventional fountain codes largely depends on $n$ and $C$ as explained in Section \ref{155959_8Apr10}. 


    The SP decoding involves mainly 4 parts, i.e.~the initialization, the check to variable computation, the variable to check computation, and the tentative decision parts. 
    Let $X$ be the random variable of a transmitted bit $x$, 
    and let $Y$ be the random variable of the corresponding channel output $y$.  
    The a posterior probability $     Q(x|y):=\Pr(X=x|Y=y),$
    for $x=0,1$ and $y\in\mathcal{A}$ is assumed to be known to the decoder, where $\mathcal{A}$ is the receiving alphabet. 
    \\\initialization:\\
    The decoders collected $n=(1+\epsilon)k/C$ channel outputs, $y_i$ for $i\in I$, where $\#I=n$. 
    Define  $I_v:=\{i\in I\mid v_i=v\}$. It follows that $I=\cup_{v=1}^NI_v$. 
    For each variable node $v$ in $\mathcal{C}_1$ for $v=1,\dotsc, N$, calculate $p_{v}^{(0)}(x)$ for $x\in \GF(2^m)$ as follows. 
    \begin{align}
    p_{v}^{(0)}(x)&= \xi\prod_{i\in I_v}\tilde{p}^{(0)}_i(h_ix)\label{181252_11Jan10}\\
    \tilde{p}^{(0)}_i(x)&=\left\{
    \begin{array}{ll}
     Q(0|y_i)&\text{if the $w_i$-th bit of $x$ is 0}\\
     Q(1|y_i)&\text{if the $w_i$-th bit of $x$ is 1},
    \end{array}
    \right.\nonumber
   \end{align} 
where $\xi$ is the normalized factor such that $\sum_{x\in\GF(2^m)}p_{v}^{(0)}(x)=1$.
    Each variable node $v=1,\dotsc, N$ in $\mathcal{C}_1$ sends the initial message $p_{vc}^{(0)}=p_{v}^{(0)}\in\R^{2^m}$ to 
    each adjacent check node $c$. 
    Set the iteration round as $\ell:=0$. 
    \\\\\ctov:\\ 
    For each check node $c=1,\dotsc,M$  in $\mathcal{C}_1$,
    let $V_c$ be the set of the adjacent variable nodes. It holds that $\#V_c=3$,   since the pre-code $\mathcal{C}_1$ is $(2,3)$-regular. 
    Each $c$ has 3 incoming messages $p_{vc}^{(\ell)}$ for $v\in V_c$ from the 3 adjacent variable nodes.
    The check node $c$ sends the following message ${p}^{(\ell+1)}_{cv}\in\R^{2^m}$ to each adjacent variable node $v\in V_c$. 
    \begin{align*}
    &\tilde{p}^{(\ell)}_{vc}(x) = {p}^{(\ell)}_{vc}(h_{vc}^{-1} x) \text{ for $x\in\GF(2^m)$},\\
    &\tilde{p}^{(\ell+1)}_{cv}= \otimes_{v'\in V_c\backslash{\{v\}}}\tilde{p}^{(\ell)}_{v'c} \\
    &{p}^{(\ell+1)}_{cv}(x) = \tilde{p}^{(\ell+1)}_{cv}(h_{vc} x)\text{ for $x\in\GF(2^m)$}.
   \end{align*} 
    where $p_1\otimes p_2\in\R^{2^m}$ is convolution of $p_1\in\R^{2^m}$ and $p_2\in\R^{2^m}$. To be precise, 
    \begin{equation*}
    (p_1\otimes p_2)(x) = \sum_{\substack{y,z\in\GF(2^m)\\x=y+z}}{p_1(y)p_2(z)} \text{ for $x\in\GF(2^m)$}.
   \end{equation*}
    The convolution seems the most complex part of the decoding. 
    Indeed, the convolutions are efficiently calculated via FFT and IFFT \cite{4155118}, \cite{RaU05/LTHC}.
    Increment the iteration round as $\ell:=\ell+1$. 
    \\
    \\\vtoc: \\
    Each variable node $v=1,\dotsc,N$ in $\mathcal{C}_1$ has 2 adjacent check nodes since the pre-code $\mathcal{C}_1$ is $(2,3)$-regular. 
    Let $C_v$ be the set of adjacent check nodes. 
    The message $p^{(\ell)}_{vc}\in\R^{2^m}$ sent from $v$ to $c\in C_v$ is given by
    \begin{align*}
    p^{(\ell)}_{vc}(x) = p_v^{(0)}(x)\prod_{c'\in C_v\backslash\{c\}}p^{(\ell)}_{c'v}(x) \text{ for $x\in\GF(2^m)$}.
   \end{align*} 
    \tentative \\
    For each $v=1,\dotsc, N$, the tentatively estimated $v$-th transmitted symbol is given as 
    \begin{align*}
    \hat{x}_v^{(\ell)}=\mathop{\mathrm{argmax}}_{x\in \GF(2^m)}\prod_{c'\in C_v}p_v^{(0)}(x)\textstyle p^{(\ell)}_{cv}(x). 
   \end{align*} 
    If $\hat{x}^{(\ell)}:=(\hat{x}_1^{(\ell)},\dotsc,\hat{x}_N^{(\ell)})$ forms a codeword of $\mathcal{C}_1$, i.e.  $\hat{x}^{(\ell)}$ satisfies every  parity-check equation of $\mathcal{C}_1$
    \begin{align*}
    \sum_{v\in V_c}h_{cv}\hat{x}_v^{(\ell)}=0\in \GF(2^m) 
   \end{align*}
    for all $c=1,\dotsc,M$, the decoder outputs $\hat{x}^{(\ell)}$ as the estimated codeword. 
    Otherwise repeat the check to variable, variable to check  and tentative steps. If the iteration round $\ell$ reaches at a pre-determined 
    number, the decoder collects more channel outputs and start over the decoding. 
    \section{Analysis of Asymptotic Overhead}
    \label{001437_10Apr10}
    In this section,  we investigate the overhead $\epsilon$ in the limit of many information bits $k\to\infty$ for the transmissions over the BEC, i.e., $C=1$. 
    Rathi developed the density evolution which enables the prediction of the decoding performance of the non-binary LDPC codes in the limit of large code length. 
The density evolution usually gives,  for a given code ensemble, the maximum channel erasure probability, referred to as threshold,  at which 
    the average decoding erasure probability goes to zero. 
    We will use the density evolution calculating the maximum overhead $\epsilon$ at which the average decoding erasure probability goes to zero in the limit of $k\to\infty$. 


   The density evolution used in this section was originally  developed for the non-binary LDPC code ensembles with parity-check matrices defined over the general linear group $\GL(\GF(2),m)$. 
   However, Rathi reported that the threshold for the code ensemble defined over $\GF(2^m)$ and $\GL(\GF(2),m)$ also have approximately the  same threshold within the order of $10^{-4}$. 
   Consequently, we shall evaluate the threshold of the proposed codes by the density evolution for $\GL(\GF(2),m)$. 

   In the binary case, we can predict the asymptotic decoding performance of LDPC codes 
   transmitted over 
   the general MBIOS channels
   in the large code length limit by {\it density evolution} \cite{richardson01design}. 
   Density evolution is possible also for the non-binary LDPC codes \cite{1273653} but computationally intensive and tractable only for the BEC. 
   The analysis for the BEC often helps us to capture the universal properties of LDPC codes. 

   When the transmission is taken place over the BEC and all-zero codewords are assumed to be sent, the messages, described by probability vectors $(p(x))_{x\in\GF(2^m)}$ of length $2^m$ in general, 
   can be reduced to linear subspaces of $\GF(2)^m$ \cite{RaU05/LTHC}. 
   To be  precise, for each message in the SP decoding algorithm, a subset of $\{\bm{x}\in \GF(2)^m\mid p(x)\neq 0\}$
   forms a linear subspace of $\GF(2)^m$, where $\bm{x}$ is the binary representation of $x\in\GF(2^m)$.

   For messages in SP decoding, probability vectors $\underline{P}=(P_0,\dotsc, P_m)$ are used for the density evolution and referred to as {\itshape densities}. 
   The $i$-th entry $P_i$ is the probability that a message forms a subspace of dimension $i$ for $i=1,\dotsc,m$. 
   Define two densities $\underline{P}^{(\ell)}$ and  $\underline{Q}^{(\ell)}$ as the densities of messages sent from variable nodes and check nodes
   at the $\ell$-th iteration round, respectively. 
   In \cite{4111/THESES}, Rathi proved that the density that outgoing messages from a variable (resp.~check) node of degree 3 with two incoming messages
   of density $\underline{P}$ and $\underline{Q}$ is given by $\underline{P}\boxdot \underline{Q}$ (reps.~$\underline{P}\boxtimes \underline{Q}$). 
   The detail calculation of $\underline{P}\boxdot \underline{Q}$ and $\underline{P}\boxtimes \underline{Q}$ are defined 
\footnote{
\vspace{-0.5cm} \begin{align*}
  &\left[ \underline{P} \boxdot \underline{Q} \right]_k = \textstyle\sum_{i=k}^m \sum_{j=k}^{k+m-i} C_{\boxdot }(m,k,i,j) P_i Q_j,\\
  &\left[ \underline{P} \boxtimes \underline{Q} \right]_k = \textstyle\sum_{i=0}^k \sum_{j=k-i}^{k} C_{\boxtimes }(m,k,i,j) P_i Q_j,\\
  &{C_\boxdot }(m,k,i,j) :=2^{(i-k)(j-k)}{\gb{i}{k}\gb {m-i} {j-k}  }\Big/{\gb m j } ,\\
  & {C_\boxtimes }(m,k,i,j) :=2^{(k-i)(k-j)} {\gb {m-i} {m-k} \gb i {k-j} }\Big/{\gb m {m-j} } ,
 \end{align*}
 where $\displaystyle\gb{m}{k}=\prod_{l=0}^{k-1}\frac{2^m-2^l}{2^k-2^l}$
 is a 2-Gaussian binomial. 
} 
in below.
   Using these 2 operations of 2 densities, the density evolution in \cite{4111/THESES} gives recursive update equations of 
   $\underline{P}^{(\ell)}$ and $\underline{Q}^{(\ell)}$ for $\ell\ge 0$.

   Rathi \cite{RaU05/LTHC} developed the density evolution for the BEC that tracks probability densities of the dimension of the linear subspaces. 
   For $\ell\ge 0$, the density evolution tracks the probability vectors $\underline{P}^{(\ell)}$ and $\underline{Q}^{(\ell)}$
    which are referred to as {\itshape densities}. 
   The initial messages in Eq.~\eqref{181252_11Jan10} can be seen as the intersection of $d$ subspaces of the messages received as the channel outputs.

   With $\epsilon$ overhead, the decoder has $k(1+\epsilon)/C$ channel outputs transmitted over the channel with capacity $C$. 
   The number of variable node in $\mathcal{C}_1$ is $N$. It holds that $N=3(N-M)=3mk$,  since $\mathcal{C}_1$ is of rate $1/3$ and defined over $\GF(2^m)$. 
   The average number of collected channel outputs per variable node in $\mathcal{C}_1$ is given by $(1+\epsilon)/(3m)=:\beta$. 
   It follows hat the probability $R_d$ that a randomly chosen variable node in $\mathcal{C}$ has $d$
   corresponding channel outputs is given by 
   \begin{align*}
    R_d&=\binom{N}{d}\left(\frac{\beta}{N}\right)^d\left(1-\frac{\beta}{N}\right)^{N-d}. 
   \end{align*}
   It follows that 
   \begin{align}
    \sum_{d\ge 0}R_dx^d&=\left(\frac{\beta}{N}x+1-\frac{\beta}{N}\right)^N\nonumber\\
    &\hspace{-3.5mm}\stackrel{(N\to\infty)}{=}e^{-\beta(1-x)}
    =\sum_{d\ge 0}\frac{\beta^de^{-\beta}}{d!}x^d.\label{192119_9Apr10}
   \end{align} 
   From this, we see the probability that a randomly chosen variable node in $\mathcal{C}$ has $d$
   corresponding channel outputs in the limit of $k\to\infty$ is $\frac{\beta^de^{-\beta}}{d!}$. 
   The density of the initial messages is given by $\underline{P}^{(0)}$ as follows,
   \begin{align*}
    &\underline{P}^{(0)}=\sum_{d\ge 0}\frac{\beta^de^{-\beta}}{d!}\overbrace{\underline{E} \boxdot\cdots\boxdot \underline{E}}^{\text{$d$ times}},
   \end{align*}
   where $\underline{E}$ is a density such that the subspace is of dimension $m-1$ with probability 1. 
   In precise, $\underline{E}:=(E_0,\dotsc, E_m),$ 
   \begin{align*}
    &E_i:=
    \left\{\begin{array}{ll}
     1 & \text{ if }i=m-1\\
    0 & \text{ if }i\neq m-1.
   \end{array}
    \right.
   \end{align*}

   Since the pre-code is a (2,3)-regular LDPC codes, we have recursive update equations of densities as follows. 
   \begin{align*}
    & \underline{Q}^{(\ell+1)}=\underline{P}^{(\ell)}\boxtimes \underline{P}^{(\ell)},\quad  \underline{P}^{(\ell+1)}=\underline{P}^{(0)}\boxdot \underline{Q}^{(\ell+1)}. 
   \end{align*}
   Since the messages of dimension 0 corresponds to the successful decoding, the asymptotic overhead $\epsilon^{\ast}$ is defined as follows. 
   \begin{align*}
    &\epsilon^{\ast}:=\sup_{\epsilon\in[0,1]}\{\epsilon\in[0,1]\mid\lim_{\ell\to\infty}P^{(\ell)}_0=1\}.
   \end{align*}
   It follows that, in the limit of  many information bits $k\to\infty$, with overhead $\epsilon<\epsilon^{\ast}$  the reliable transmissions are possible with the proposed $C_\infty$. 


   \begin{table}
       \caption{Asymptotic overhead $\epsilon^{\ast}$ of the proposed codes $\mathcal{C}_\infty$ with a pre-code $(2,d_\mathrm{c})$-regular non-binary LDPC codes over  $\GF(2^m)$ transmitted over the BEC, i.e.,  $C=1.0$.  }
       \label{041650_16Apr10}
    \begin{center}
{\scriptsize
     \begin{tabular}{|l|c|c|c|c|}
      \hline
      $m$&$d_\mathrm{c}=3$&$d_\mathrm{c}=4$&$d_\mathrm{c}=5$&$d_\mathrm{c}=6$\\\hline\hline
      1&1.0799&3.3945&5.9311&8.6557\\\hline
      2&0.5748&2.3274&4.2477&6.3098\\\hline
      3&0.3295&1.8033&3.4128&5.1370\\\hline
      4&0.2075&1.5341&2.9732&4.5078\\\hline
      5&0.1422&1.3816&2.7151&4.1293\\\hline
      6&0.1069&1.2910&2.5536&3.8855\\\hline
      7&0.0888&1.2359&2.4487&3.7210\\\hline
      8&0.0809&1.2025&2.3786&3.6068\\\hline
      9&\bf{0.0792}&1.1826&2.3312&3.5256\\\hline
      10&0.0813&1.1716&2.2987&3.4665\\\hline
      11&0.0856&1.1661&2.2765&3.4228\\\hline
      12&0.0913&\bf{1.1645}&2.2613&3.3904\\\hline
      13&0.0977&1.1653&2.2511&3.3659\\\hline
      14&0.1044&1.1677&2.2445&3.3472\\\hline
      15&0.1111&1.1713&2.2405&3.3331\\\hline
      16&0.1179&1.1754&2.2383&3.3222\\\hline
      17&0.1245&1.1801&\bf{2.2378}&3.3142\\\hline
      18&0.1309&1.1851&2.2380&3.3081\\\hline
      19&0.1371&1.1901&2.2392&3.3036\\\hline
     \end{tabular}
}
    \end{center}   
   \end{table}
   Table~\ref{041650_16Apr10} shows the asymptotic overhead $\epsilon^{\ast}$ of the proposed code $\mathcal{C}_\infty$ over $\GF(2^m)$ for different $m=1,\dotsc, 19$. 
   Table~\ref{041650_16Apr10} also lists the asymptotic overheads with (2,$d_\mathrm{c}$)-regular non-binary LDPC pre-code for $d_\mathrm{c}$=4, 5 and 6.
   It can be seen that the best overhead $\epsilon^\ast=0.079$ is attained at $d_\mathrm{c}=3$ and $m=9$ and
   the fountain code $\mathcal{C}_\infty$ exhibit  very poor overhead if defined on $\GF(2^m)$ with $m=1$, i.e,. the binary field. 
   We will use $m=8$, for its good asymptotic overhead $\epsilon^{\ast}=0.081$ in Tab.~\ref{041650_16Apr10} and friendliness for byte-oriented processors. 
    \section{Numerical Results}
\begin{figure}
    \begin{center}
     \includegraphics[width=8cm]{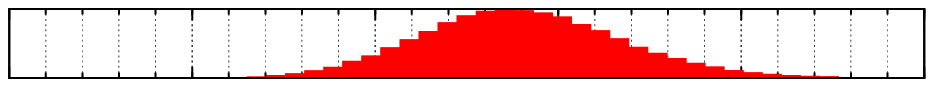}\vspace*{-4mm}
     \includegraphics[width=8cm]{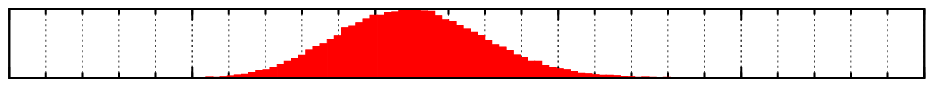}\vspace*{-4mm}
     \includegraphics[width=8cm]{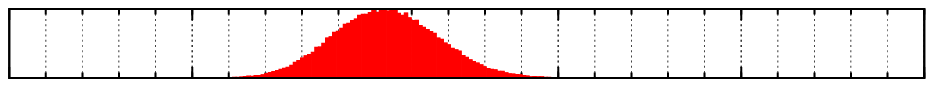}\vspace*{-4mm}
     \includegraphics[width=8cm]{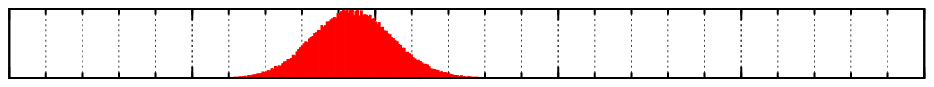}\vspace*{-4mm}
     \includegraphics[width=8cm]{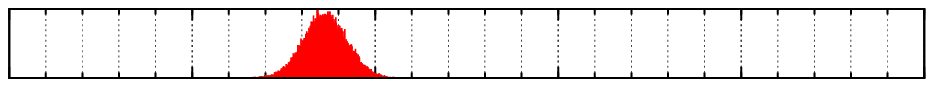}\vspace*{-4mm}
     \includegraphics[width=8cm]{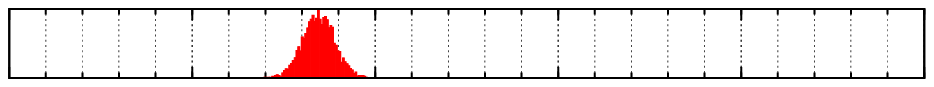}\vspace*{-4mm}
     \includegraphics[width=8cm]{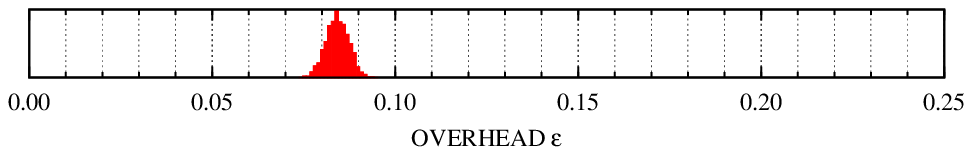}
     \caption{Histograms of the overheads at which the proposed fountain code over $\GF(2^8)$ successfully recovers  $k$ information bits over the channel with $C=1.0$. 
     The number of the information bits is set $k=192, 512, 1024, 2048, 8192, 16384$ and $32768$ from the top to bottom. 
     The horizontal axis describes the overhead $\epsilon$. 
     It can be seen that it is getting concentrated at overhead 0.081 as predicted in Tab.~\ref{041650_16Apr10} at $m=8$. 
}
     \label{234703_9Apr10}
    \end{center}
   \end{figure}

\begin{figure}
    \includegraphics[scale=1.0]{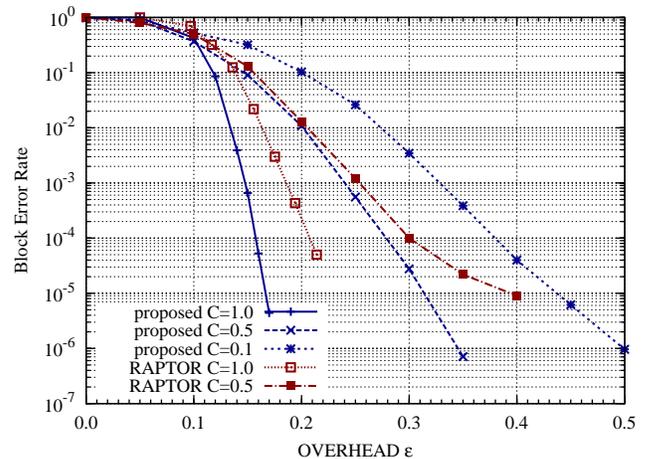}
    \caption{Decoding performance of the proposed fountain codes for the binary-input AWGN channels with capacity $C$=1.0, 0.5 and 0.1, The information size is $k=1024$. 
    The performance of best-so-far Raptor codes \cite{VPD08}, \cite{Ven08}, \cite{1702289} optimized for $k=1024$  are drawn for comparison. }
    \label{234653_9Apr10}
   \end{figure}

    In this section, we present demonstrations of $\mathcal{C}_\infty$ defined over $\GF(2^8)$ with small and moderate information bits. 
    Transmission over the BEC and the AWGN channels are considered.   
    Fig.~\ref{234703_9Apr10} shows the histograms of overheads of $\mathcal{C}_\infty$ defined over $\GF(2^8)$. It seems that the asymptotic overhead is getting concentrated 
    at 0.081 as predicted in Section \ref{001437_10Apr10}. 
    Fig.~\ref{234653_9Apr10} shows the decoding performance of the proposed fountain code transmitted over the binary-input AWGN channels with capacity $C=1.0, 0.5$ and $0.1$. 
    The horizontal axis describes the overhead and the vertical axis describes the block error rate. 
    The proposed codes exhibit the better performance than the best-so-far Raptor codes. 
   \section{Conclusion}
   In this paper we propose a new simple fountain coding scheme whose decoding complexity does not depend on the number of collected channel outputs.
   No optimization of the output degree distribution is needed. Because of the non-binary property, we believe the proposed codes can be used for memory channel.
    \bibliographystyle{IEEEtran}
     \bibliography{IEEEabrv,kenta_bib}
   \end{document}